\documentclass[journal]{IEEEtran}
\usepackage{multirow}
\usepackage{overpic}
\usepackage{color}
\usepackage{bm}
\usepackage{cite}

\ifCLASSINFOpdf
\else
\fi
\usepackage{fixltx2e}
\begin{document}
%
\title{Silicon photomultiplier-based Compton Telescope \\
for Safety and Security (SCoTSS)}
%
%
%
\author{Laurel~Sinclair,
        Patrick~Saull,
        David~Hanna,
        Henry~Seywerd,
        Audrey~MacLeod and
        Patrick~Boyle
\thanks{L.E.~Sinclair, H.C.J.~Seywerd and A.M.L.~MacLeod are with the
  Geological Survey of Canada, Natural Resources Canada, 601~Booth~St, Ottawa,
Ontario, K1A 0E8, Canada. e-mail: laurel.sinclair@nrcan.gc.ca.}
\thanks{P.R.B.~Saull is with Measurement Science and Standards, National
  Research Council, 1200~Montreal~Rd, Ottawa, Ontario, K1A 0R6, Canada.}
\thanks{D.S.~Hanna is with the Physics Department, McGill University, 3600~University~St, Montreal, Quebec, H3A 2T8, Canada.}
\thanks{Manuscript received Dec 20, 2013.}}

%
%

\markboth{Full article appears in IEEE Trans.Nucl.Sci. 61:5 2745 - 2752, October~2014}%
{Shell \MakeLowercase{\textit{et al.}}: Bare Demo of IEEEtran.cls for Journals}
\maketitle
%

\begin{abstract}
A Compton gamma imager has been developed for use in consequence management
operations and in security investigations. The imager uses solid
inorganic scintillator, known for robust performance in field survey
conditions.  The design was constrained in overall size by the requirement that it be person transportable and
operable from a variety of platforms. In
order to introduce minimal dead material in the path of the incoming and
scattered gamma rays, custom silicon photomultipliers (SiPMs), with
a thin glass substrate, were used to collect the scintillation light from the
scatter layers. To move them out of the path of the gamma rays,
preamplification electronics for the silicon photomultipliers were located a
distance from the imager.  This imager, the Silicon photomultiplier Compton 
Telescope for Safety and Security (SCoTSS) is able to provide a one-degree image
resolution in a $\pm$45$^{\circ}$ field of view for a 10~mCi point source 40 m
distant, within about one minute, for gamma-ray energies ranging from 344~keV
to 1274~keV.  Here, we present a comprehensive performance study of the SCoTSS imager.
\end{abstract}

\begin{IEEEkeywords}
Compton imager, gamma imager, Compton telescope, gamma camera, Compton camera, SiPM, silicon photomultiplier.
\end{IEEEkeywords}

%
\IEEEpeerreviewmaketitle

\section{Introduction}
%
%
%
%
\IEEEPARstart{C}{anadian} federal
radiological assessment team partners, as well as border security operators,
have years of experience using large-volume inorganic scintillators to detect
radioactive substances from a distance via their gamma emissions, in outdoor
mobile survey conditions~\cite{Sinclair20111018,IAEA_1995,IAEA_1991,Morning_Light}.
These groups would benefit enormously from the use of a mobile device capable of providing
an image of a radioactive substance overlaid on a photograph
of the field of view.  
Operated in a non-imaging total-count mode, an imager having a sensitive volume of the order of several 
litres would provide the high
sensitivity and fast detection times that mobile survey crews are used to, with the option
of switching to imaging mode once a radiation field of interest has been detected.

The imaging modality would bring the obvious benefit of providing a faster
source localization than could be provided by a non-directional detector operating
in a raster pattern.  An imager is capable of localizing multiple or distributed 
sources, which would give omnidirectional detectors a problem.  Additionally, it could allow
for localization of a source without a potentially hostile party under investigation
being aware of it.
Alternately, it would allow investigation of a scene without 
disturbing it, providing personnel safety from loose sources of radioactivity, and 
preserving evidence for eventual prosecution.  

Some earlier work in developing imagers for safety and security made use of solid-state devices such as high purity germanium~\cite{Si_LLNL_2006}.  For field operation, however, it would be best to avoid the use of instruments which require cryogenic cooling.  Progress has been made to produce a Compton imager using room temperature solid-state devices~\cite{Polaris_2004}, and also there are other teams developing Compton imagers using scintillator~\cite{Lee2010,Jung2013}.  These, however, do not have the sensitivity we require.  Notable parallel efforts have taken place to develop highly sensitive fieldable gamma imagers~\cite{Ziock_2008,SORDS_2009,MISTI_2009}.  These are not, however, suitable to our application as their very large size and weight necessitates operation from a particular dedicated survey platform.  

The Compton gamma-imaging technique is optimal for mobile methods, in that as
opposed to coded-aperture or pinhole imaging, it does not require the
transport of a heavy gamma-ray shield.
Moreover, the recent development of very small and low-mass silicon
photomultipliers 
(SiPMs)~\cite{SiPM_review_2011,SiPM_rev_2006,SiPM_rev_2004}
permits the collection of light from within a layered Compton gamma imager,
with negligible scattering induced in the dead material of the light
collection device~\cite{us_2012}.

We have designed and built an imager for use by nuclear emergency response and
security teams, using solid inorganic scintillator with SiPM
light collection.
This paper presents the first performance results for this detector, the
Silicon photomultiplier Compton Telescope for Safety and Security (SCoTSS).

\section{Compton imaging}

\begin{figure}
  \centering
  \includegraphics[height=5.5cm]{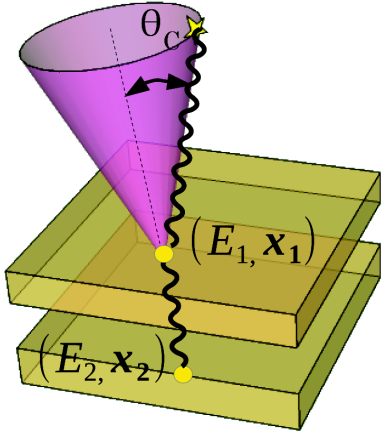}
  \caption
  { \label{fig:imager_diagram} 
    Generic Compton imager showing a Compton scatter with energy $E_1$ at position $\bm{x_1}$ in the scatter detector, and the subsequent absorption of the scattered gamma ray leading to energy deposit $E_2$ at position $\bm{x_2}$ in the absorber detector.  The cone-shaped locus of possible source locations as reconstructed from the energy deposits is also shown.
  }   
\end{figure}

A Compton imager relies on measurement of the energy and position of a Compton scatter, $E_1$ and $\bm{x_1}$, and on the measurement of the energy and position of the absorption of the scattered gamma ray, $E_2$ and $\bm{x_2}$, in order to reconstruct the origin of an incoming gamma ray~\cite{Nature_1974}.  A generic Compton imager, as illustrated in 
Fig.~\ref{fig:imager_diagram}, is thus composed of two parts, a scatter detector at the front, in which the Compton scatter should occur, and an absorber detector behind that, to collect the scattered gamma ray.
For simple events consisting of a single Compton scatter followed by photoabsorption of the scattered gamma ray, the scatter angle, $\theta_C$, is given by the expression~\cite{Compton_1923}
\begin{equation}
\cos \theta_C = 1 + \mbox{m$_{\mbox{\scriptsize e}}$}c^2(1/E_{\mbox{\scriptsize tot}} - 1/E_2),
\end{equation}
where $E_{\mbox{\scriptsize tot}} = E_1 + E_2$, $\mbox{m$_{\mbox{\scriptsize e}}$}c^2$ is the electron rest mass, and the approximation has been made that the electron from which the incoming gamma ray scatters is unbound and at rest.
Thus, the position of the emitter can be reconstructed up to an arbitrary azimuthal angle, to lie on the surface of a cone with axis along the line $\bm{x_1} - \bm{x_2}$, and opening angle $\theta_C$.
By overlaying several such cones, the location or distribution of the source of gamma rays may be determined.
Note that the image resolution of a Compton imager is tied intrinsically to the imager's energy resolution -- meaning that a good Compton imager is also a good spectroscopic instrument which may be used to identify the isotopic composition of sources which are a priori unknown.


\section{The imager}

   \begin{figure*}[h!]
   \begin{center}
   \begin{tabular}{c}
   \begin{overpic}[height=5.5cm]{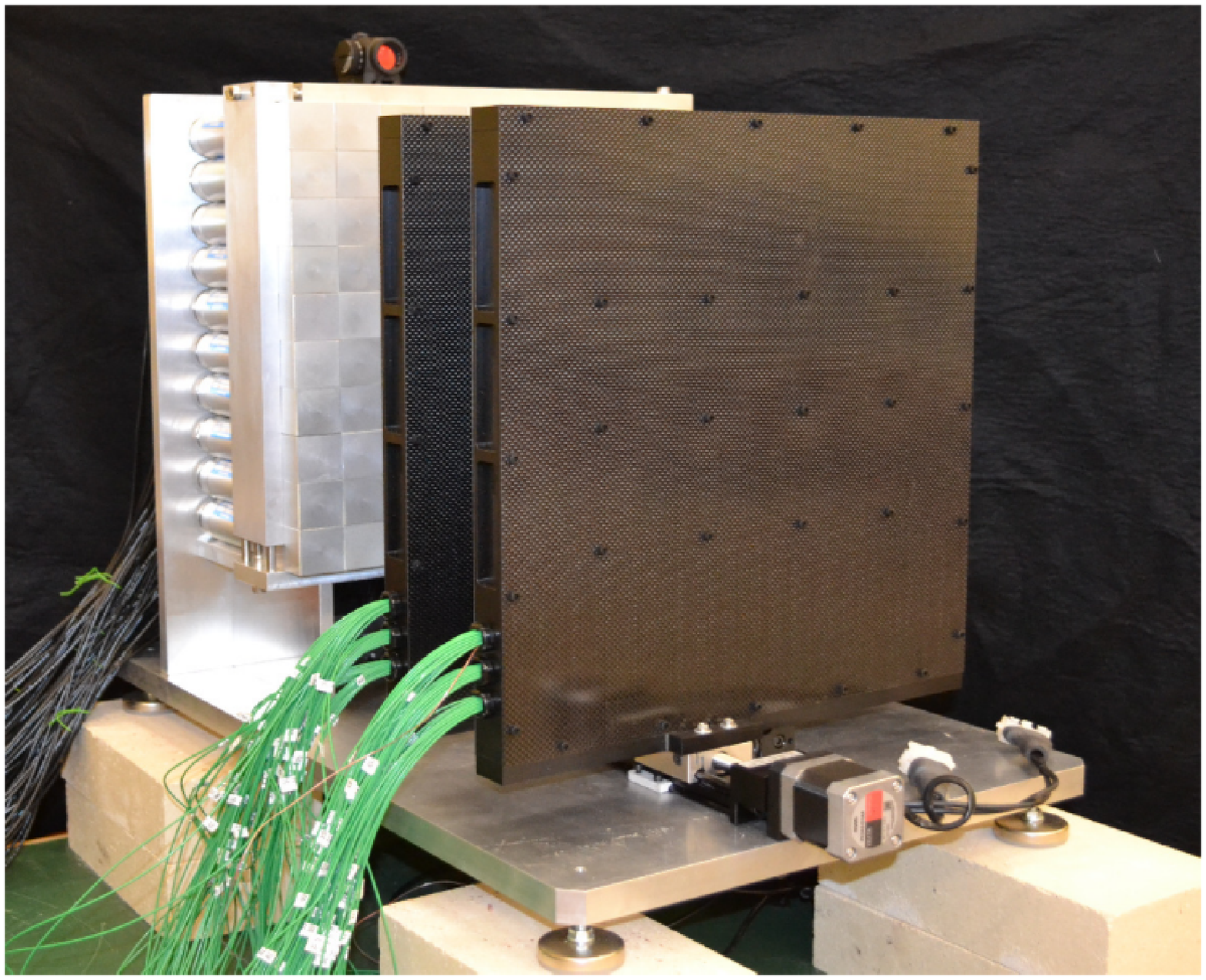}\put(6,70){\textcolor{white}{a)}}\end{overpic}
   \hspace{0.7cm}
   \begin{overpic}[height=5.5cm]{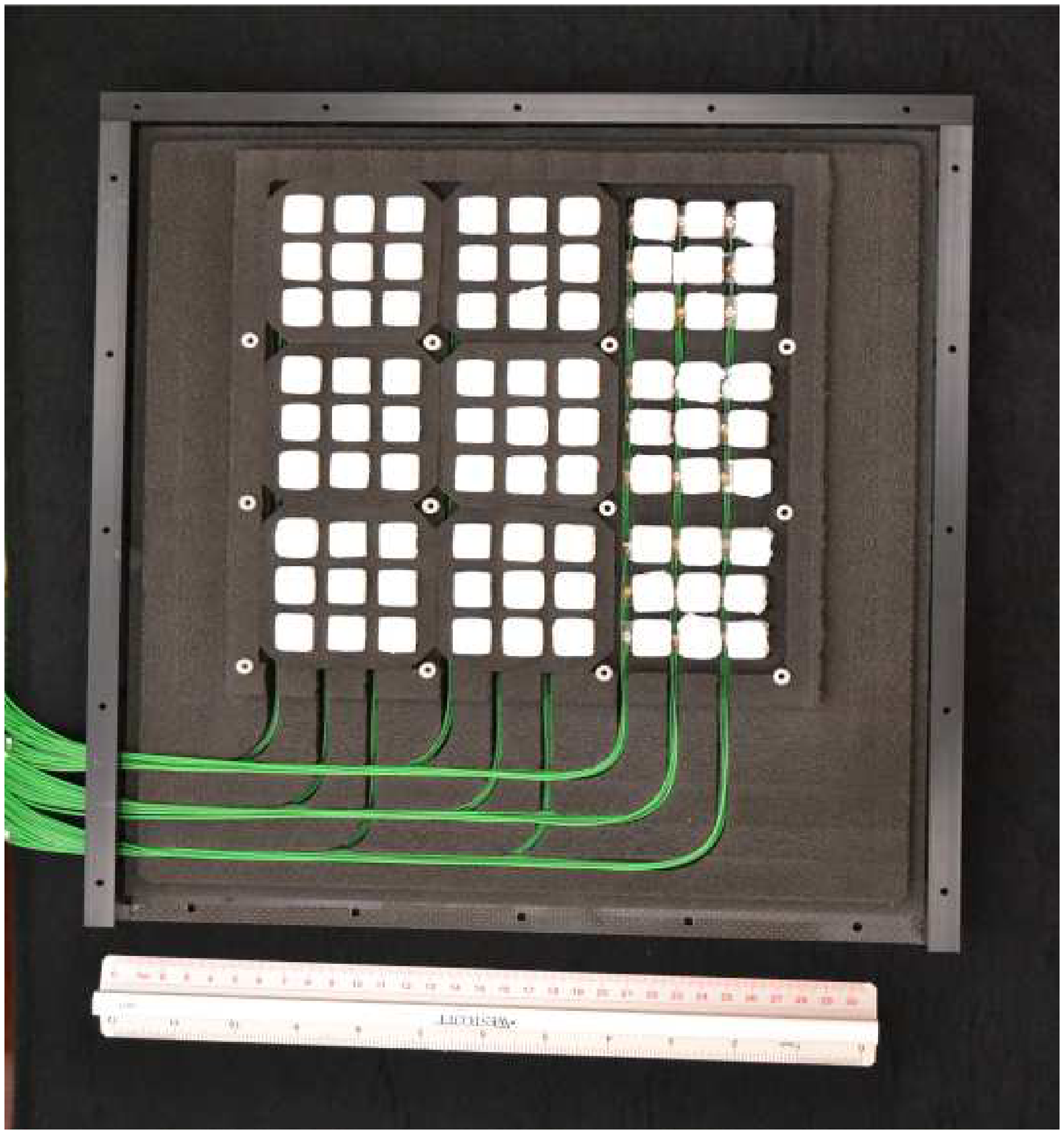}\put(8,85){\textcolor{white}{b)}}\end{overpic}\\
   \vspace{0.4cm}\\
    \begin{overpic}[height=4.2cm]{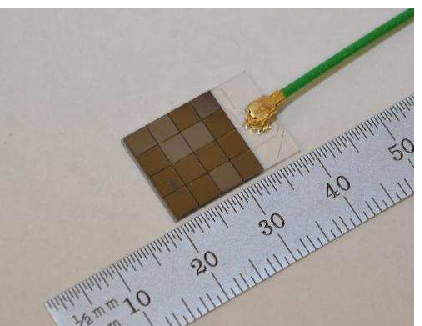}\put(8,65){\textcolor{white}{c)}}\end{overpic}
   \hspace{1.6cm}
   \begin{overpic}[height=4.2cm]{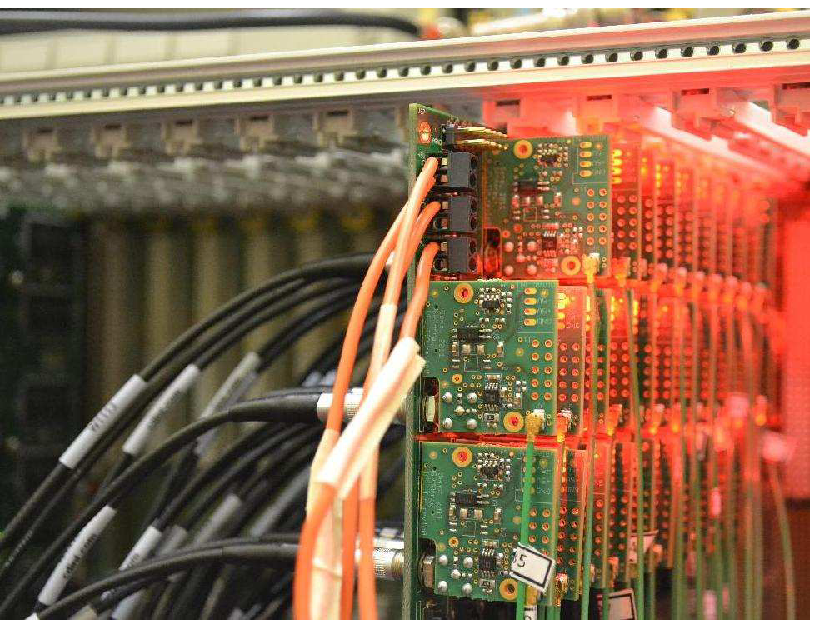}\put(8,65){\textcolor{black}{d)}}\end{overpic}
   \end{tabular}
   \end{center}
   \caption
   { \label{fig:the_imager} 
a) The imager with the
two scatter layers on the right and the absorber detector behind them at the
left.  The 10 x 10 array of aluminum-clad PMT~+~NaI(Tl) assemblies is visible at the rear of the imager, clamped in an aluminum frame.  The two scatter layers are shown enclosed in their carbon-fibre and Delrin frame, with micro-coax cables exiting the layers to the left.  
The Velmex translators to which the scatter layers are affixed for adjusting
the inter-layer spacing, can be seen underneath the scatter layers.
A ``red dot'' alignment sight is visible on top of the absorber detector.
b) The inside of a scatter layer showing the eighty-one CsI(Tl) crystals
wrapped with thread-seal (``Teflon'') tape.  Foam spacers have been removed from the three
right-most columns of pixels to show the micro coax cables and connections to
the glass of the SiPMs.  
c) A single SiPM detector unit used for reading out a single CsI(Tl) crystal.  d) The bias distribution and pre-amplification boards for the
SiPM~+~CsI(Tl) channels, connected to the motherboard.  
   }
   \end{figure*} 

An extensive design phase has been conducted to determine the optimal imager
materials and geometry~\cite{twopixel_2010,us_2009}.  We considered the following
parameters:

\subsubsection{Total Size}
 
The imager total weight and exterior volume were constrained by the
  requirement that teams composed of one or two people should be able to lift
  the imager from a transport container into a survey vehicle or platform,
  without machine assistance.  The imager, pictured in Fig.~\ref{fig:the_imager} a), was thus constrained to occupy a
  cubic volume approximately 35~cm on a side, with a total mass of about 25~kg.

\subsubsection{Sensitive Volume}

  Within the constraints on the exterior dimensions, the
  imager should have as large a sensitive volume as possible.  This
  increases sensitivity to weak sources.  Our scatter
  detector consists of two 9~x~9 layers of cubic pixels of CsI(Tl) 1.35~cm on a side.  This
  crystal arrangement in an open scatter layer is shown in
  Fig.~\ref{fig:the_imager} b).
  The
  absorber detector consists of a 10~x~10 layer of
  2.5~x~2.5~x~4.0~cm$^3$ pixels of NaI(Tl).  Thus, the total scintillator volume of
  the imager is 2\,898~cm$^3$.  This is comparable to the total scintillator
  volume of current standard omnidirectional mobile survey systems (10~x~10~x~41~cm$^3$).
  Thus, second-by-second the imager provides the level of minimum
  detectable activity to which nuclear emergency response teams are currently accustomed, even without 
  using it in an imaging mode.

\subsubsection{Light Collection}

  We chose to use SiPMs for light collection in the scatter layers because the
  SiPMs could be made extremely thin and low mass such that they would present
  minimal dead material to the Compton-scattered gamma ray.  The SiPMs were
  manufactured by SensL~\cite{SensL} on a
  glass substrate only 0.5~mm thick.  An individual SiPM module, used to read out a single
  CsI(Tl) crystal, is shown in
  Fig.~\ref{fig:the_imager} c).  The SiPM module consists of
  16 pads, a total of 76\,384 Geiger-mode
  avalanche photodiodes, which are summed to provide a single analogue signal from
  the 1.35~x~1.35~cm$^2$ active area of the SiPM.
  The bias supply and pre-amplification circuits were located on
  front-end electronics boards a distance from the gamma imager and
  therefore out of the path of the incident and scattered gamma rays 
  (see Fig.~\ref{fig:the_imager} d)).

  We used conventional photomultiplier tubes (PMTs) for the
  light collection from the pixels of the absorber detector.  SiPMs could have
  been used for the absorber detector as well, but dead material is
  not a problem at the back of the absorber detector, and at the time of
  parts procurement for the prototype imager, PMTs were better proven and less 
  expensive.  The individual aluminum-clad NaI(Tl)~+~PMT assemblies are clearly
  visible in the rear plane in Fig.~\ref{fig:the_imager} a). 

\subsubsection{Scintillating Material}

  Despite the large initial-state electron effects on image resolution of high-Z
  scintillator materials, materials like NaI(Tl), CsI(Tl) and LaBr$_3$(Ce)
  still perform better than organic scintillator due to their much superior energy
  resolution~\cite{us_2009}.  The SensL SiPM which we use has a peak light
  sensitivity at 480~nm, and therefore CsI(Tl), with its relatively high
  peak wavelength of emission, was chosen for the material of
  the scatter detector.  The mobile survey workhorse, NaI(Tl), which is well matched in
  peak wavelength of light emission to standard PMTs, was selected for the
  rear detector.  Our aim has been to develop a
  prototype for eventual commercialization and adoption by nuclear emergency
  response teams and therefore, although an earlier study had determined that its 
  performance was optimal~\cite{us_2009}, LaBr$_3$(Ce) was ruled out for
  either component due to price constraints.

\subsubsection{Detector Thickness}

  As the thickness of the scatter detector increases, the probability for a
  Compton scatter to be initiated increases, but at the same time the
  probability for a second scatter, or photoabsorption event, also
  increases.  We have optimized the total thickness of the scatter detector by
  choosing the thickness which maximizes the probability for an incoming gamma ray of energy around 1~MeV to scatter once, and
  then exit the scatter detector, according to the approach described in ~\cite{us_2009}.

  To obtain a better measurement of the depth of the position of the Compton scatter
  within the total thickness of the scatter detector, we have segmented the
  scatter detector into two layers.  These may be seen in
  Fig.~\ref{fig:the_imager}.
  In this work, events were selected with one energy deposit in the scatter
  detector, in either layer.  In the future, events featuring
  Compton scatters in both scatter layers, and also low energy events where
  the photoabsorption process occurs in the second scatter layer, could be utilized.

  For the absorber detector, efficiency increases with increasing thickness,
  but since we do not have a depth measurement in the absorber detector, there
  is increasing uncertainty on the depth of the energy
  deposit with increasing crystal depth,
  leading to a loss of image resolution.  We used EGSnrc~\cite{EGSnrc1,EGSnrc2} simulation studies to
  balance the conflicting requirements of high absorption probability and good
  position precision by choosing the depth of the absorber detector which gives 
  the desired image precision in the shortest time~\cite{us_2009}.

\subsubsection{Pixel Size}

  A smaller lateral physical dimension of the pixels in both the scatter
  and absorber detectors would provide a better measurement of the positions of
  the energy deposits and therefore improved measurement of the cone axes and, ultimately, improved image
  resolution.  However, making the pixels too small would increase the cost due to additional readout channels.
  We have optimized the performance versus cost by choosing the lateral pixelization to
  contribute approximately the same image smearing as is contributed by
  initial-state electron effects and finite energy resolution, following the approach described in~\cite{us_2009}.

\subsubsection{Layer Spacing}

  Increasing the separation between the scatter detector and the absorber detector in a Compton gamma
  imager will result in improved angular resolution, at a cost of lowered efficiency and narrowed field of view.
  We chose not to set the inter-layer distance at a fixed separation, optimized
  for a particular application.
  Instead, both scatter layers of our imager are affixed to Velmex
  XSlides~\cite{Velmex} to allow for varying the inter-layer distance according to
  experimental or operational need.  Monte Carlo simulation studies determined that for the large source distances
  under study here, the time to achieve an image of a certain precision dropped as the distance between the rear scatter layer 
  and the absorber detector was increased, up to about 14~cm, beyond which no significant improvement or worsening was seen.  
  For the measurements 
  presented here, the two scatter layers are centred laterally with respect to the absorber detector, and the distances between the centres 
  of the scatter layers  and the front face of the absorber layer are fixed at 
  14.6~cm and 23.9~cm.

\subsubsection{Electronics}

  For the purposes of detector performance optimization and characterization, power supply and readout of the
  imager has been accomplished with standard multi-purpose crate-based electronics.  For the imager to be eventually
  fieldable from a variety of platforms, custom electronics with a suitable small and rugged form will have to be developed.

In the final configuration for this study, the detector features two scatter
layers, each with a 9~x~9 arrangement of cubic pixels of CsI(Tl), 
1.35~cm on a side, as
shown in Fig.~\ref{fig:the_imager} b).  The CsI(Tl) crystals are
connected to the SiPM with NyoGel OC-462 optical gel~\cite{Nye}, and then this unit is wrapped in thread-seal 
tape (commonly referred to as Teflon tape).
Within a layer, the
pixels are grouped into nine 3~x~3 arrangements, to allow for cable routing.
The distance from the centre of one pixel to the next within one 3~x~3 group
is 2~cm.  The distance from the centre of one 3~x~3 group to the next is
7~cm.  Mechanical support for the scatter layers is provided by low-density
foam on 0.8~mm-thick carbon-fibre sheets mounted in a Delrin frame.
Custom front-end boards, shown in Fig.~\ref{fig:the_imager} d), provide both
the SiPM bias voltage and pre-amplification.  Further details are available 
in~\cite{us_2012}.

The 2.5~x~2.5~x~4~cm$^3$ NaI(Tl) crystals of the absorber detector are
read out with Hamamatsu~\cite{Hamamatsu} R1924A PMTs.  Proteus~\cite{Proteus}
provided custom-designed assemblies with outer dimension 3.2~x~3.2~x~15.2~cm$^3$,
as small as possible in order to minimize the dead space between the individual NaI(Tl) crystals.
An aluminum frame provides compression to pack the NaI(Tl) + PMT assemblies
tightly in a 10~x~10 array.  The high voltage of the PMTs is supplied using a
CAEN~\cite{CAEN} HV 1535SN unit and the PMT high
voltages are trimmed such that each channel reconstructs the 662 keV photopeak of
$^{137}$Cs at the same pulse height.

\section{Data acquisition}

We investigated the performance of the imager by placing a $^{137}$Cs point source of
known strength at various angles $\theta$ with respect to the symmetry axis of the
detector.
We used several different isotopes, $^{152}$Eu, $^{22}$Na,
$^{137}$Cs, and $^{54}$Mn, and were thus able
to investigate the detector performance as a function of gamma-ray energy as
well.
The detector has been configured for optimal performance at long range, and
therefore where possible the sources were placed at a distance of 10~m from the centre of
the front face of the detector.  The $^{152}$Eu and $^{54}$Mn
sources were weaker so they were placed at 5~m from the centre of the front
face.  (In the investigation in Section~\ref{subsec:TTI} of how long it takes to achieve an image of a
desired precision, we scale the
results such that all of the sources are effectively of the same strength at
the same distance.)  The run configurations for the data sets presented in
this study are shown in Table~\ref{tab:runs}.
\begin{table*}[!h]
  \centering
  \caption{  \label{tab:runs}
Data sets}
  \begin{tabular}{ c c c c c c c c c}
    \hline
\vspace*{-.2cm}\\
    \multirow{3}{*}{Run}&\multicolumn{4}{ c }{Source} &\multirow{3}{*}{Distance}&\multirow{3}{*}{$\theta$}&\multirow{3}{*}{Live time}&\multirow{3}{*}{Number of triggered events}\\
    \cline{2-5}                                            

\vspace*{-.2cm}\\
               &Isotope         &Energy  & Emission probability    &Rate           &       &                     &         &  \\
               &                &(keV)   & (\%)  &(s$^{-1}$)    &  (m)  & ($^\circ$)          & (s)     &  \\
    \hline                                                                     
\vspace*{-.2cm}\\
    1          & $^{137}$Cs      & 662  &  85.0 &   $2.1 \times 10 ^{7}$            &10.0   & 0.0                 & 11236 &  $1.8 \times 10^{7}$ \\
                                                         
    2          & $^{137}$Cs      & 662  &  85.0 &     $2.1 \times 10^{7}$          &10.0   & 10.0                & 11218 &  $1.8 \times 10^{7}$  \\
                                                            
    3          & $^{137}$Cs     & 662   &  85.0 &      $2.1 \times 10^{7}$        &10.0   & 20.0                & 11119 &  $1.8 \times 10^{7}$ \\
                                                            
    4          & $^{137}$Cs      & 662  &  85.0 &     $2.1 \times 10^{7}$         &10.0   & 30.0                & 10968 &  $1.8 \times 10^{7}$ \\
                                                            
    5          & $^{137}$Cs      & 662  &  85.0 &      $2.1 \times 10^{7}$        &10.0   & 40.0                & 11067 &  $1.8 \times 10^{7}$ \\
                                                            
    6          & $^{152}$Eu    & 344   &  27.7  &      $1.6\times 10^{6}$       &5.0    &20.0                 & 17817 & $1.2\times 10^{8} $\\
                                                            
    7          & $^{22}$Na    & 511   &  180.8   &  $2.6 \times 10^{7}$          &10.0    & 20.0               & 10524 & $ 2.3 \times 10^{7}$ \\
                                                            
    8          & $^{54}$Mn     & 835   &  100.0  &   $4.3 \times 10^{6}$          &5.0    & 20.0                & 40667 &  $6.0 \times 10^{7}$  \\
                                                            
    7          & $^{22}$Na     & 1\,274   &  100.0  &  $1.6 \times 10^{7}$        &10.0    & 20.0               & 10524 & $ 2.3 \times 10^{7}$ \\
    \hline
  \end{tabular}
  \\
\end{table*}

\begin{figure}[h!]
  \centering
  \includegraphics[height=5.2cm]{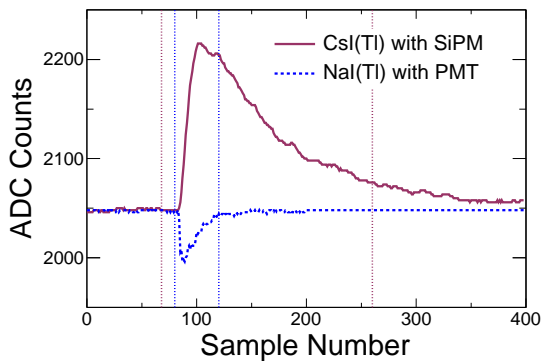}
  \caption{
    \label{fig:coincident_pulse} 
    Coincident traces, digitized with a 16~ns sampling rate, from a scatter and an absorber pixel.  Vertical
    lines show the integration windows.}
\end{figure} 
The signals from all channels are digitized every 16~ns using CAEN V1740 64-channel digitizers, with two boards for the PMTs and three for the SiPMs.
Fig.~\ref{fig:coincident_pulse} shows coincident pulses from the scatter and absorber detector, as read out by the digitizers, where it can be seen that CsI(Tl) has a much longer decay time than NaI(Tl).
Triggering is performed by the CAEN digitizers in conjunction with a CAEN V1495 general purpose VME logic board.
We trigger solely on the rise of the pulses in the absorber detector (NaI(Tl)~+~PMT) channels.  Whenever a NaI(Tl)~+~PMT channel sample crosses a user-defined ADC threshold, a local trigger is issued.  An OR of these local triggers is provided at the front panel of each of the two PMT V1740 boards as a NIM signal.
These two signals are fed to the V1495, where they are ORed and fanned out to the external trigger inputs of all five boards in the DAQ system. Thus, the external trigger initiates the global storage of an event in VME memory across all five  boards. The board clocks are carefully synchronized so that the event timestamps for coincident data agree. These timestamps are monitored during data taking to ensure that the boards do not fall out of synchronization (not observed during any of the runs described here). Note that the local trigger threshold can only be set for groups of eight channels on the V1740 which, in conjunction with a residual channel-to-channel baseline variation, leads to an effective spread in the trigger threshold across all channels of about $\sim$20~keV.  The trigger threshold for data taking corresponds to a pulse energy of approximately 200~keV for all runs except $^{152}$Eu, where it was lowered to about 50~keV.

To keep the data rate within the VME system limit (40 MB/s), the pulse samples were integrated on board the V1740 digitizers using custom firmware supplied by CAEN, and only the pulse areas read out.  For each channel on a board, the firmware provides a simple sum of the digitized samples within a desired integration window. The user defines this window by specifying the offset time with respect to the trigger and the number of 
samples over which to integrate. These parameters are common to all channels on a board, so it was necessary to split the readout into two boards of NaI(Tl)~+~PMT and three boards of CsI(Tl)~+~SiPM channels to account for the 
different decay times. The firmware also provides a method for channel-by-channel online baseline subtraction, so that the resulting sample sums are proportional to pulse area and therefore energy. 

The pulse integration window for the CsI(Tl)~+~SiPM channels is chosen to coincide with where a coincident pulse would fall with respect to the PMT pulse (see Fig.~\ref{fig:coincident_pulse}). Consequently, although we do not trigger directly on the CsI(Tl)~+~SiPM channels, if there are pulses from these boards coincident with the PMT trigger their energies are correctly reconstructed.  

For each triggered event, sample-sums from all five boards are read out to Linux workstation memory.  To limit the amount of data
written to disk, the sample-sums from only those channels with energy 
greater than three standard-deviations of the pedestal (typically 10~keV for PMT energies and 25~keV for the SiPM energies) are
written to disk for further offline processing. Our trigger rate for naturally occuring background radiation, for the 200~keV trigger threshold, is about 1100~Hz.  This increased to at most 2200~Hz for runs with a source present, corresponding to at most 1\% dead time.  For the $^{152}$Eu run at lower threshold, the trigger rate increased to 6700 Hz (3\% dead-time). No effects due to pulse pile-up were observed in the data.

Note that our trigger does not require a coincidence in time between the rising edges of the scatter and absorber pulses.  Instead, whatever energy is reconstructed in the scatter layer integration window is considered to be due to a coincident pulse, unless the energy deposit fails the threshold cut, or the event fails the selection cuts.  This is done intentionally, to maintain the sensitivity of our detector when it is functioning in a gross-count (non-imaging) mode as a survey spectrometer.

Energy calibration of the detector is carried out using a variety of gamma sources with known energies. For the scatter detector, the energy scale is parametrized linearly using the 662~keV peak of $^{137}$Cs and the zero-energy pedestal.
For the absorber detector, the energy scale is parametrized with a third-degree polynomial using emissions over the range 40~keV to 2614~keV from $^{241}$Am, $^{113}$Sn, $^{152}$Eu, $^{22}$Na, $^{137}$Cs, $^{40}$K and $^{208}$Tl.  We achieve mean energy resolutions of 7.9\% (29\%) and 7.5\% (23\%) at 662~keV (60~keV) for CsI(Tl)~+~SiPM and NaI(Tl)~+~PMT channels, respectively.

Data taking was carried out after the setup had been warmed up over several days in a lab having stable temperature. We found that drifts in the response of the CsI(Tl)~+~SiPM channels could be limited to about 1\% through daily re-calibrations with a $^{137}$Cs source. For the NaI(Tl)~+~PMT channels, a high voltage trim on a monthly basis was sufficient to accomplish the same. 

\section{Event Selection}
\label{sec:selection}

\begin{figure*}[h!]
  \begin{center}
    \begin{tabular}{c}
      \begin{overpic}[height=5.2cm]{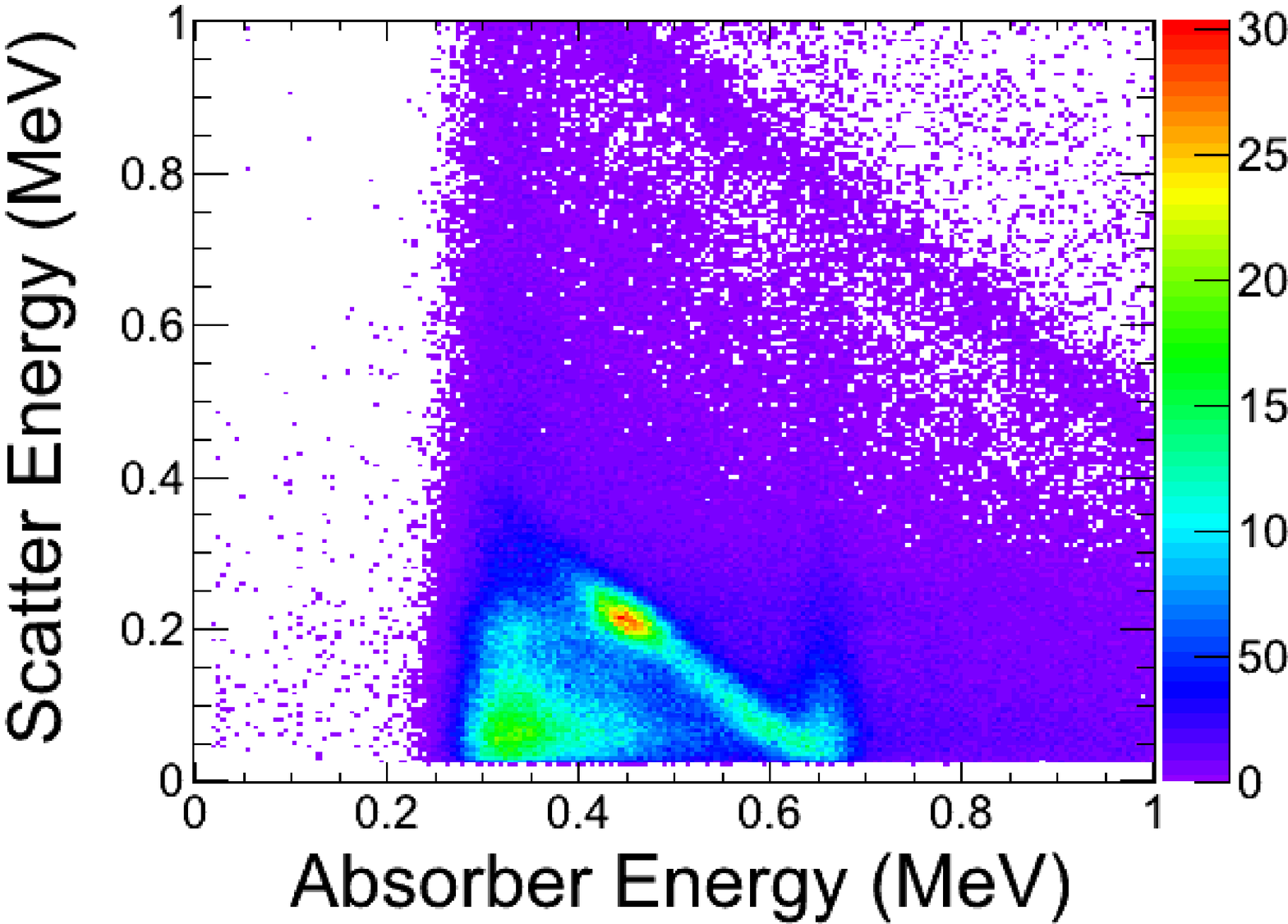}\put(8,65){\textcolor{black}{a)}}\end{overpic}
      \hspace{0.5cm}
      \begin{overpic}[height=5.2cm]{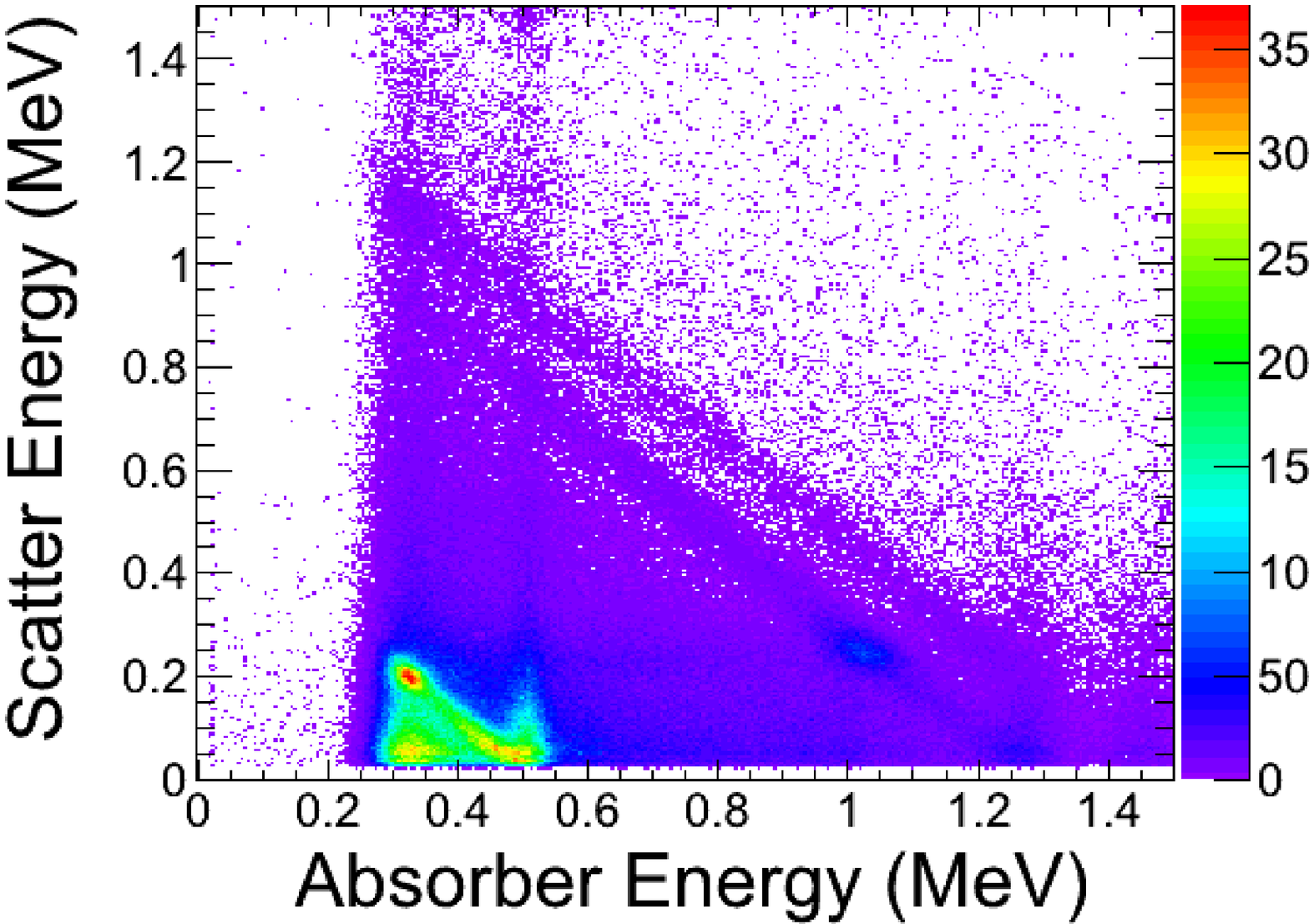}\put(8,65){\textcolor{black}{b)}}\end{overpic}
    \end{tabular}
  \end{center}
  \caption
      { \label{fig:selection} 
        a) Scatter energy versus absorber energy for Run 3, $^{137}$Cs source at 20$^{\circ}$.
        b) Scatter energy versus absorber energy for Run 7, $^{22}$Na source at 20$^{\circ}$. 
        }
\end{figure*} 
Only events with exactly one energy deposit in each of the scatter and absorber detectors were considered for further analysis. 
The measured scatter energy, E$_{\mbox{\scriptsize scat}}$, is shown versus the measured absorber energy, E$_{\mbox{\scriptsize abs}}$, for two-hit events for Run 3, $^{137}$Cs at 20$^\circ$, in Fig.~\ref{fig:selection}a) and for Run 7, $^{22}$Na at 20$^\circ$, in Fig.~\ref{fig:selection}b).
Full-energy deposition events with  E$_{\mbox{\scriptsize scat}}$ and E$_{\mbox{\scriptsize abs}}$ summing to the photopeak energies are clearly seen as diagonal lines with intercepts at 662~keV for $^{137}$Cs and at 511~keV and 1\,274~keV for $^{22}$Na.
 For image reconstruction, we select events where the energy sum of the two hits is within about two standard deviations of the photopeak energy of interest; see Table~\ref{tab:EvSel}.
\begin{table*}[h!]

  \centering
  \caption{\label{tab:EvSel}Event Selection}
  \begin{tabular}{ c c c l c c }
    \hline
\vspace*{-.2cm}\\
    Run        &Id             & E$_{\mbox{\scriptsize tot}}$ (keV) & Photopeak selection & Back-scatter rejection\\

    \hline                              
\vspace*{-.2cm}\\
    1 -- 5      & $^{137}$Cs  & 662                    &  $|$E$_{\mbox{\scriptsize scat}}$ + E$_{\mbox{\scriptsize abs}}$ - E$_{\mbox{\scriptsize tot}}| < $40~keV     &   E$_{\mbox{\scriptsize scat}} < 170$~keV     \\
                                                
    6           & $^{152}$Eu  & 344                     &  $|$E$_{\mbox{\scriptsize scat}}$ + E$_{\mbox{\scriptsize abs}}$ - E$_{\mbox{\scriptsize tot}}| < $33~keV       & E$_{\mbox{\scriptsize scat}} < 110$~keV       \\
                                                
    7           & $^{22}$Na   & 511                     &  $|$E$_{\mbox{\scriptsize scat}}$ + E$_{\mbox{\scriptsize abs}}$ - E$_{\mbox{\scriptsize tot}}| < $40~keV        &   E$_{\mbox{\scriptsize scat}} < 160$~keV   \\
                                                
    8          & $^{54}$Mn    & 835                      &  $|$E$_{\mbox{\scriptsize scat}}$ + E$_{\mbox{\scriptsize abs}}$ - E$_{\mbox{\scriptsize tot}}| < $45~keV        &  E$_{\mbox{\scriptsize scat}} < 170$~keV   \\
                                                
    7          & $^{22}$Na    &1\,274                     &  $|$E$_{\mbox{\scriptsize scat}}$ + E$_{\mbox{\scriptsize abs}}$ - E$_{\mbox{\scriptsize tot}}| < $80~keV       &  E$_{\mbox{\scriptsize scat}} < 170$~keV or 280~keV $<$ E$_{\mbox{\scriptsize scat}} < 635$~keV     \\
    \hline
  \end{tabular}
  \\
\end{table*}

A class of full energy deposition events, the back-scatter events (scatter in the absorber layer, absorption in the scatter layer), is apparent in both plots, tightly clustered at particular values of E$_{\mbox{\scriptsize scat}}$ and E$_{\mbox{\scriptsize abs}}$.  For example, for the $^{137}$Cs run, the full energy deposition back-scatter events all occur at E$_{\mbox{\scriptsize scat}} \sim 200$~keV and E$_{\mbox{\scriptsize abs}} \sim 460$~keV.  This kinematic tendency makes the rejection of back-scatter events straightforward.  Isotope-dependent energy requirements were applied, as shown in Table~\ref{tab:EvSel}.

\section{Results}
\label{sec:results}

\subsection{Efficiency and Angular Resolution}
\label{subsec:EfficARM}
We define efficiency as the number of events which pass the event
selection minus the number of naturally occuring background events, divided
by the total number of gamma rays crossing the 20~x~20~cm$^2$ area of the
first scatter layer during the run.  The efficiency measurements are presented
in Table~\ref{tab:ARMeff}.  The dominant uncertainties in the measurement of
efficiency are the uncertainty on the measurement of the source
emission rate and the uncertainty on the background subtraction. The efficiency does not vary strongly with angle, and varies slightly with energy, with the detector being more efficient at lower energies.
\begin{table}[!t]

  \centering
  \caption{\label{tab:ARMeff}Angular Resolution Measure and Efficiency}
  \begin{tabular}{ c c c c c l c c }
    \hline
\vspace*{-.2cm}\\
    Run        & Id           & Energy [keV] & $\theta$ ($^\circ$) & ARM ($^\circ$)& Efficiency (\%)\\
                             
    \hline                                                                     
\vspace*{-.2cm}\\
    1          & $^{137}$Cs      & 662        & 0.0      & $3.5 \pm 0.1$     & $0.59 \pm 0.11$         \\
                                  
    2          & $^{137}$Cs      & 662        & 10.0     & $3.6 \pm 0.1$     & $0.62 \pm 0.11$         \\
                                  
    3          & $^{137}$Cs     & 662         & 20.0     & $3.5 \pm 0.1$     & $0.65 \pm 0.12$  \\
                                  
    4          & $^{137}$Cs      & 662        & 30.0     & $3.5 \pm 0.1$     & $0.62 \pm 0.11$         \\
                                  
    5          & $^{137}$Cs      & 662        & 40.0     & $3.6 \pm 0.1$     & $0.61 \pm  0.11$         \\
                                  
    6          & $^{152}$Eu    & 344        &20.0    & $4.7 \pm 0.1$       & $0.78 \pm 0.21$       \\
                                  
    7          & $^{22}$Na    & 511          & 20.0   & $4.0 \pm 0.1$        & $0.70 \pm 0.13$      \\
                                  
    8          & $^{54}$Mn     & 835        & 20.0   & $3.1 \pm 0.1$       & $0.41 \pm 0.06$    \\
                                  
    7          & $^{22}$Na     & 1\,274         & 20.0     & $2.8  \pm 0.1$    & $0.40 \pm 0.07$      \\
    \hline
  \end{tabular}
  \\
\end{table}

\begin{figure}
  \centering
  \includegraphics[height=5.2cm]{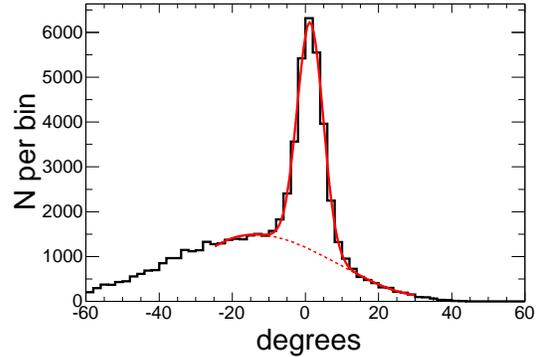}
  \caption{
    \label{fig:oneARM} 
    Angular resolution measure for Run 3, $^{137}$Cs source at 20$^{\circ}$, shown as the solid black histogram.
    A fit to the data of the sum of a Gaussian and a third-degree polynomial is shown as the solid curve.  The third-degree polynomial component is shown as the dashed curve.
    The Gaussian fit has a sigma of 3.5$^\circ$.}
\end{figure}

The Angular Resolution Measure (ARM) is defined as 
$\theta^{\mbox{\scriptsize R}}_{\mbox{\scriptsize C}} -
\theta_{\mbox{\scriptsize geom}}$, where $\theta^{\mbox{\scriptsize
    R}}_{\mbox{\scriptsize C}}$ is the Compton scatter angle as it is
reconstructed from the measured energy deposits and $\theta_{\mbox{\scriptsize
    geom}}$ is the angle between the line connecting the two energy deposits, and
the line between the first energy deposit and the source~\cite{COMPTEL_1993}.
The ARM distribution, for Run 3 with the $^{137}$Cs source at 20$^{\circ}$, is
shown in Fig.~\ref{fig:oneARM}.
The distribution shows two components.  The well-reconstructed forward-going scatters fall in the
Gaussian peaked part of the distribution. There is also a broad pedestal extending to 
negative values due to contributions from incompletely contained background radiation, random
overlap between different events, poorly-reconsructed events, and back-scatter events that pass the selection cuts.

The ARM distributions for all runs have been fit to the sum of a Gaussian distribution and a third-degree polynomial, as indicated for Run~3 in Fig.~\ref{fig:oneARM}.  The sigma values from the Gaussian fits are presented in Table~\ref{tab:ARMeff}.  We find that the
angular resolution of the imager improves with increasing energy, 
ranging from about 4.7$^{\circ}$ to 2.8$^{\circ}$.  

Efficiency and angular resolution are quality measures which can be used to compare the performance of different designs.  However, it often happens that varying a certain design parameter can be beneficial to efficiency, and harmful to angular resolution, or vice versa.  Therefore, to optimize our design, and to compare it with other designs, we make use of a quantity called ``time to image'', defined in the next section.

\subsection{Time to image}
\label{subsec:TTI}

A simple back-projection of the Compton cones into space can provide an image of the emitter with angular precision commensurate with the angular resolution.  However, this method fails to make the best use of the available statistics and of the differing uncertainties affecting each back-projected cone.  We have developed a $\chi^2$ minimization algorithm, based on MINUIT~\cite{MINUIT}, for finding the source direction which best satisfies all the Compton cones in a sample of $N$ events, taking into account the uncertainties on cone opening angle and axis. The algorithm is iterative and allows for the rejection of events with poorly fitting cones, such as those associated with naturally occuring background events, back-scatter events, events which contain a noise hit, or events with partial energy deposit. Details can be found in~\cite{twopixel_2010}. Here we summarise the algorithm used for the current analysis, which differs slightly from the version in that reference.  We note that this algorithm functions to locate a point source, providing a quantitative measure of the detector's performance, which we have used to optimise our design.   For operational use, we intend to implement other imaging modalities which will be applicable for multiple or extended sources. 

For the first iteration, a starting seed direction for the fit is found by back-projecting all cones onto a 2-D histogram with $4^\circ \times 4^\circ$ binning and selecting the direction corresponding to the most populated bin.
All events in the sample which have cone surfaces coming within five standard deviations of angular separation from this seed direction are accepted in the
  $\chi^2$ minimization step, which generates a new direction. In the second iteration, all cones in the full sample of $N$ events coming within two standard deviations of the fitted direction are considered in the fit and a new direction is generated. The third iteration is a repeat of the second, to allow more cones to be included in this final fit using the updated direction.

\begin{figure*}[htb]
  \begin{center}
    \begin{tabular}{c}
      \begin{overpic}[height=5.2cm]{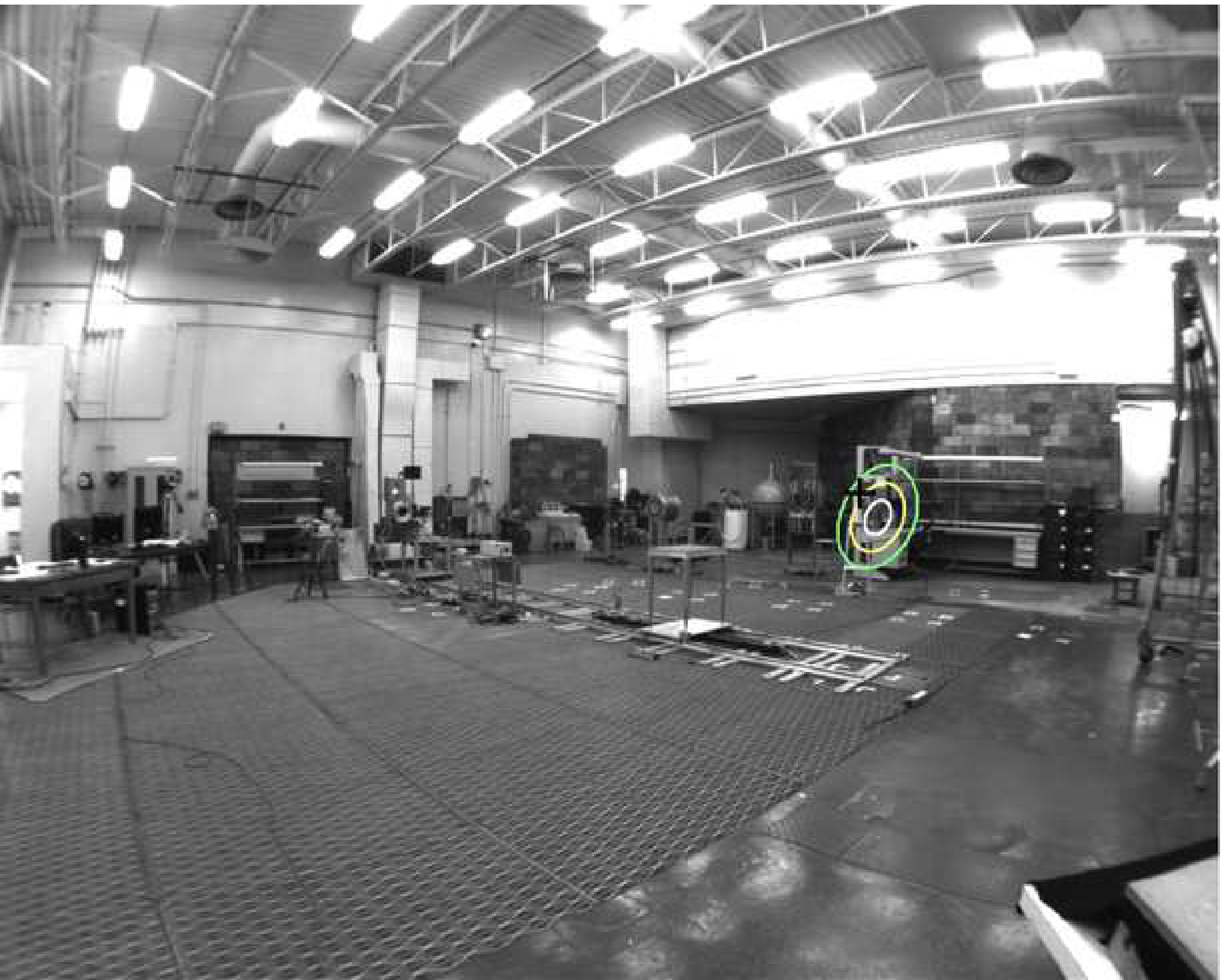}\put(5,72){\textcolor{white}{a)}}\end{overpic}
      \hspace{0.5cm}
      \begin{overpic}[height=5.2cm]{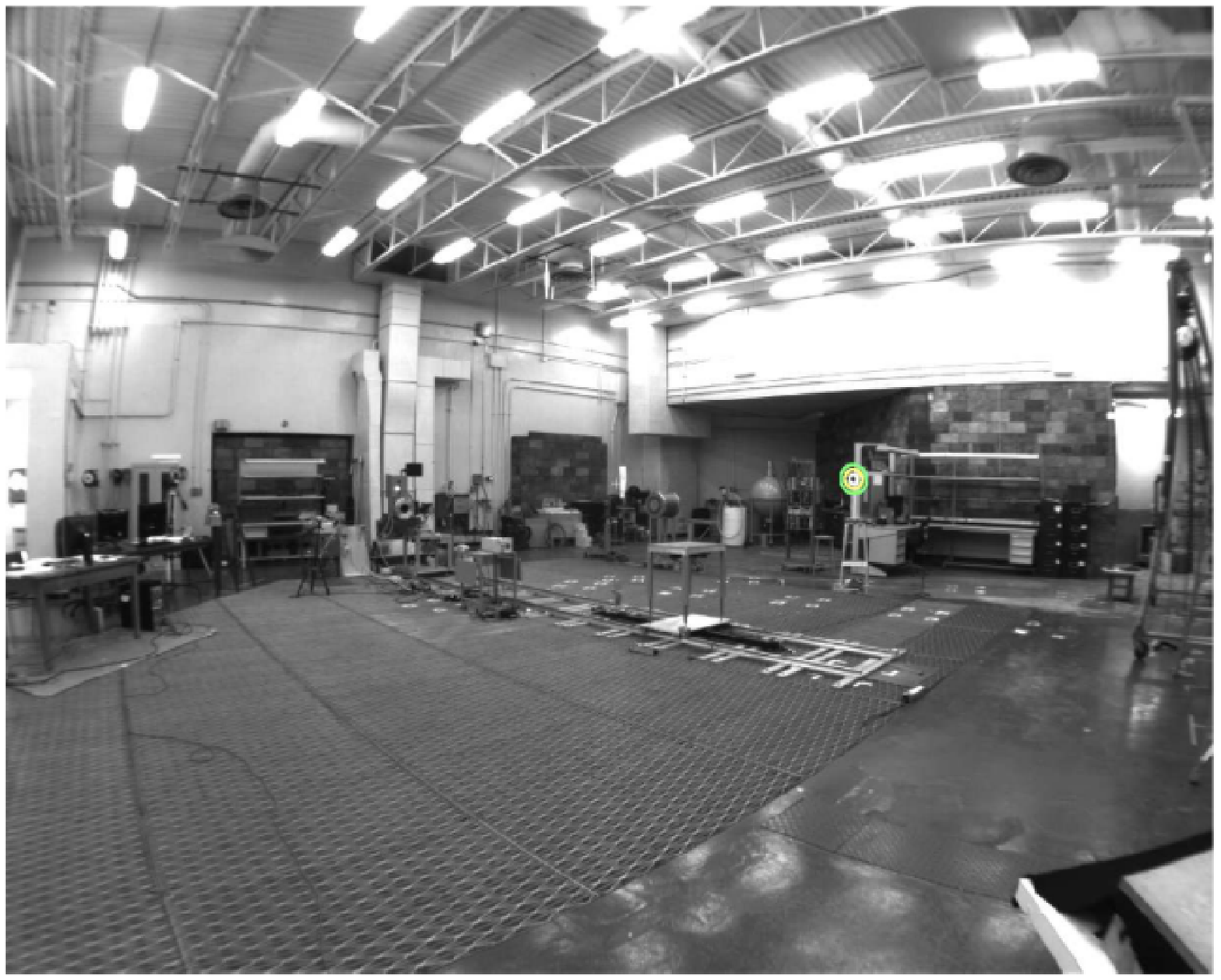}\put(5,72){\textcolor{white}{b)}}\end{overpic}
    \end{tabular}
  \end{center}
  \caption
      { \label{fig:images} 
        a) Image reconstructed using four seconds of data from Run 3, $^{137}$Cs source at 20$^{\circ}$.
        b) Image reconstructed using 60 seconds of data from Run 3, $^{137}$Cs source at 20$^{\circ}$.  The lines indicate the one-, two- and three-sigma confidence intervals.  The true source location is indicated in the left figure by a black cross.
        }
\end{figure*}

Fig.~\ref{fig:images} a) shows an image reconstructed in this way, for four seconds of data from Run 3 with the $^{137}$Cs source at 20$^\circ$.  The one-, two- and three-sigma confidence intervals are indicated by the coloured contours.
Fig.~\ref{fig:images} b) shows the image reconstructed from the same run, using 60~seconds of data.  The localization of the gamma emitter is much more precise.

\begin{figure}
  \centering
  \includegraphics[height=5.2cm]{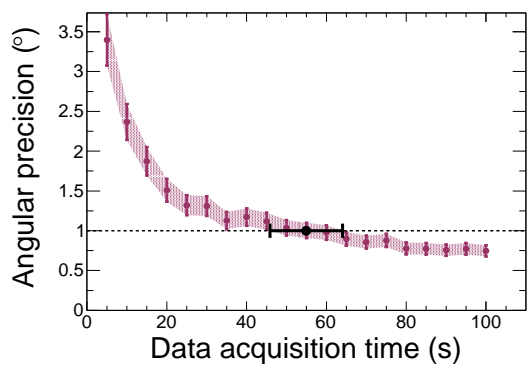}
  \caption{
    \label{fig:theta_rms} 
    Precision versus acquisition time for Run 3, $^{137}$Cs source at 20$^{\circ}$.  The acquisition time required for the precision to drop to one degree is indicated by the filled dot with horizontal error bars.}
\end{figure}

To quantify the improvement in precision with time, we look at the root mean square spread of reconstructed image directions among datasets of a certain acquisition time, for different acquisition times.  This image precision is shown versus acquisition time in Fig.~\ref{fig:theta_rms}, where the results have been scaled to reflect a 10~mCi source with 100\% branching ratio, at 40~m.
We find that angular precision improves with time, going approximately as $1/\sqrt t $ where $t$ is the aquisition time, as expected from counting statistics.  The angular precision for this configuration reaches one degree in just under one minute -- indicated in the graph by a filled dot with horizontal error bars.

\begin{figure*}[htb]
  \begin{center}
    \begin{tabular}{c}
      \begin{overpic}[height=5.2cm]{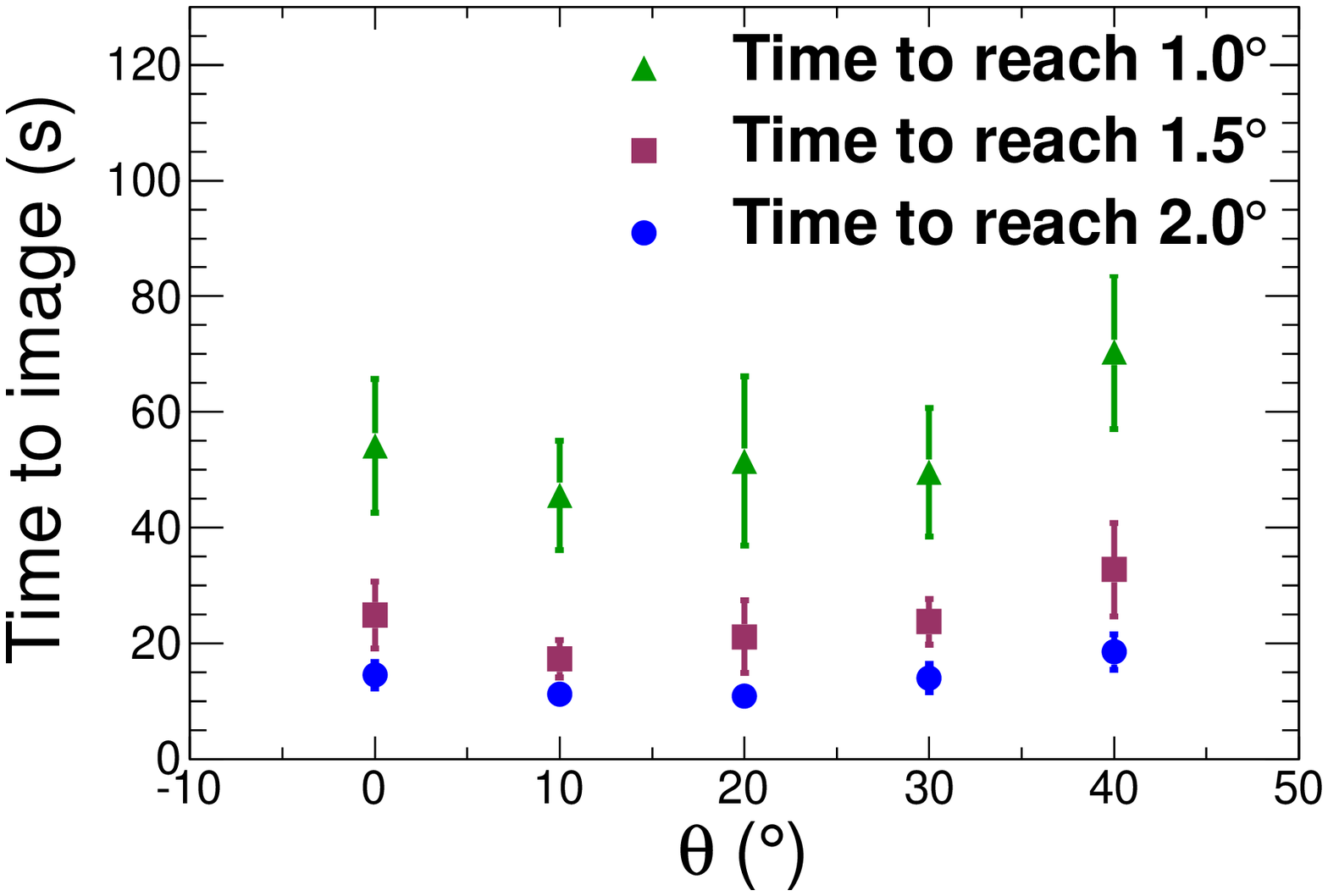}\put(18,53){\textcolor{black}{a)}}\end{overpic}
      \hspace{0.5cm}
      \begin{overpic}[height=5.2cm]{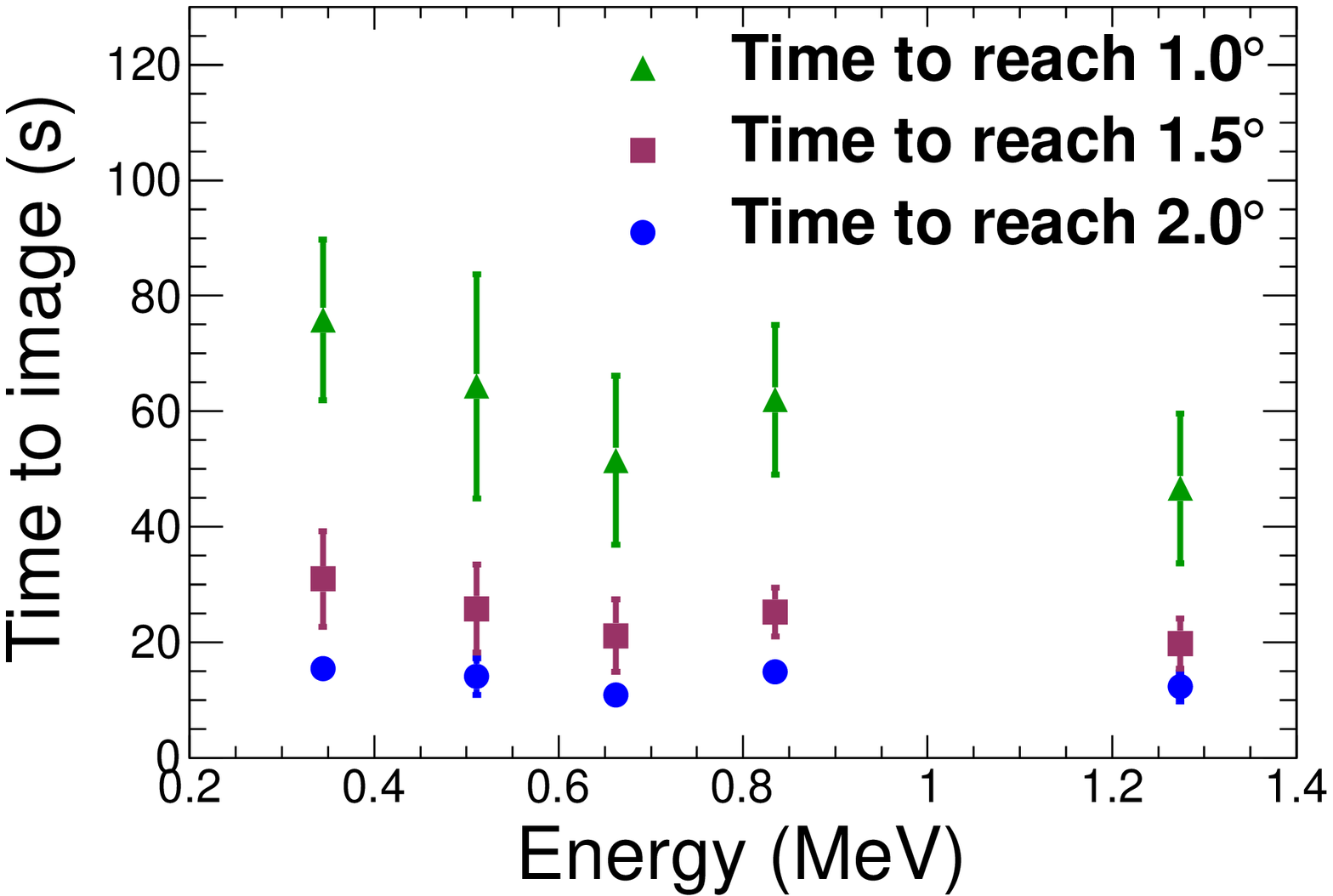}\put(18,53){\textcolor{black}{b)}}\end{overpic}
    \end{tabular}
  \end{center}
  \caption
      { \label{fig:TTI} 
        a) Time to image versus the angle between the line joining the centre of the front face of the detector and the source, and the symmetry axis of the detector.  This is for the 662~keV line of the $^{137}$Cs source and results have been scaled to represent a 10~mCi source with 100\% branching ratio at 40~m.
        b) Time to image versus energy for a source at 20$^\circ$ from the symmetry axis of the detector.  Results have been scaled to represent 10~mCi sources of 100\% branching ratio at 40~m.
        }
\end{figure*}

By taking the intercept of the precision versus acquisition time plot with horizontal lines of various desired precision values, we can look at how time to image varies for different experimental configurations.
In Fig.~\ref{fig:TTI} a) we show time to image versus the angle between the line joining the centre of the detector's front face and the source location, and the symmetry axis of the detector.
We find that generally the imager is able to achieve one degree of image precision within about a minute, across a $\pm$~45$^\circ$ field of view.  Many features of interest at 40~m will be separated by two degrees or more, for example the distance between two windows of a building, as viewed from a road.  If this is the desired precision, the imager is able to produce a useful result in less than 20~seconds, across the entire field of view.

Fig.~\ref{fig:TTI} b) shows the times to image versus photopeak energy, for a source 20$^\circ$ off-axis, where the measurements have been scaled to correspond to an isotope of 10~mCi strength, with 100\% branching ratio.  We find that the imager is able to reach one degree of precision within about a minute to a minute and a half, depending on the isotope energy.  The imager geometry is optimized for higher-energy sources, but nevertheless performs adequately down to 344~keV.  Future work could investigate the imager response to lower or higher energies of photopeak emission -- or to a highly shielded source producing an indistinct photopeak.  Times to image of better than 20~seconds are achieved across the energy range from 344~keV to 1\,274~keV for an image precision of two degrees which will be adequate for many operational circumstances.

\section{Conclusions}
A Compton gamma imager for use in nuclear security and emergency response, has been designed, built and characterized.
The imager is based on solid inorganic scintillator, which has proven performance in outdoor conditions, and read out with custom-made low-mass silicon photomultipliers.  The imager has an intrinsic angular resolution of about four degrees.  Image precisions of one degree in a $\pm$45$^\circ$ field of view can be obtained for a 10~mCi point source at 40~m within about one minute, for gamma-ray energies ranging from 344~keV to 1\,274~keV.  Two-degree image precisions are obtainable within 20~seconds.

This achievement has been obtained with early generation SiPMs.  Further improvements in efficiency are to be expected with future improvements in SiPM technology as noise levels are reduced and it becomes possible to lower energy thresholds.  
A future, fieldable version of this instrument would have to be environmentally isolated and make use of the individual bias settings featured on the SiPMs to control gain fluctuations due to temperature.
Warm up time for the instrument would likely be around 15 minutes and a price point in the low 100~k~US\$'s should be feasible.
The total volume of scintillator and corresponding second-by-second isotope identification ability will allow the fieldable version of this instrument to function as a drop-in replacement for the current ubiquitous non-directional NaI(Tl) mobile survey spectrometers.

\section*{Acknowledgments}

This work was supported by Canada's Centre for Security Science, project CRTI
07-0193RD.  The authors thank A.W.~Saull for assistance with graphics.
This is NRCan/ESS Contribution number 20130419.

\ifCLASSOPTIONcaptionsoff
  \newpage
\fi



\bibliographystyle{IEEEtran}
\bibliography{IEEEabrv,GI_SiPM}

\begin{thebibliography}{10}
\bibitem{CAEN}
{Digitizers and high voltage supplies are available from CAEN S.p.A.,
  Viareggio, Italy}.
\end{thebibliography}
%

\end{document}